\documentclass[
superscriptaddress,
preprint,
 amsmath,amssymb,
 aps
]{revtex4-2}

\usepackage{graphicx}
\usepackage{color}
\usepackage{bm}

\begin{document}

\author{Haolin~Li}
\affiliation{Department of Physics and Astronomy and Manchester Centre for Nonlinear Dynamics, University of Manchester, Oxford Road, Manchester M13 9PL, UK}
\author{Anne~Juel}
\affiliation{Department of Physics and Astronomy and Manchester Centre for Nonlinear Dynamics, University of Manchester, Oxford Road, Manchester M13 9PL, UK}
\author{Finn~Box}
\affiliation{Department of Physics and Astronomy and Manchester Centre for Nonlinear Dynamics, University of Manchester, Oxford Road, Manchester M13 9PL, UK}
\author{Draga~Pihler-Puzovi\'{c}}
\affiliation{Department of Physics and Astronomy and Manchester Centre for Nonlinear Dynamics, University of Manchester, Oxford Road, Manchester M13 9PL, UK}
%\aff{1}\corresp{\email{draga.pihler-puzovic@manchester.ac.uk}}}

\title{The propagation of air fingers into an elastic branching network}

\begin{abstract}
 We study experimentally the propagation of an air finger through the Y-bifurcation of an elastic, liquid-filled Hele-Shaw channel, as a benchtop model of airway reopening. With channel compliance provided by an elastic upper boundary, we can impose collapsed channel configurations into which we inject air with constant volume-flux. We typically observe steady finger propagation in the main channel, which is lost ahead of the Y-bifurcation but subsequently recovered in the daughter channels. At low levels of initial collapse, steady finger shapes and bubble pressure in the daughter channels map onto those in the main channel, despite small differences in initial collapse in different parts of the Y-channel. However, at higher levels of initial collapse where the elastic sheet almost touches the bottom boundary of the channel, experimentally indistinguishable fingers in the main channel can lead to multiple states of reopening of the daughter channels. The downstream distance at which steady propagation is recovered in the daughter channels also varies considerably with injection flow rate and initial collapse because of a transition in the mechanics regulating finger propagation. We find that the characteristic time and length-scales of this recovery are largest in the regime where viscous and surface tension forces dominate at low flow rate and/or low initial collapse, and that they decrease towards a constant plateau reached in the limit where elastic and surface tension forces balance at high flow rate and/or high initial collapse. Our findings suggest that practical networks are unlikely to comprise long enough channels for steady state propagation to remain established.
\end{abstract}

\pacs{??}

\maketitle

\section{Introduction}
\label{sec:intro}

Networks of elastic channels containing low Reynolds number multiphase flows are abundant in the human body. Compliance of these channels and ensuing interaction between fluids and elastic walls is crucial for the physiology of airways \citep{AlencarNature2002} and small blood vessels \citep{Basetti2016}, for example. Often these vessels are collapsed, as in the case of bronchioles before pulmonary airway reopening, so that fluid flows in narrow gaps bounded by elastic walls \citep{tePas2008}. In microfluidics, harnessing the interaction between deformable features (e.g. elastic walls) and fluid flows offers a route to passive  lab-on-chip devices that do not rely on off-chip hardware \citep{Stone2009}. However, studies of channel networks with elastic components are still in their infancy. 

Here we present an experimental study of two-phase flow through a Y-bifurcation in a rigid rectangular channel with an elastic upper boundary. Before the start of the experiment, the channel is filled with liquid which is then drained to induce deformation of the upper boundary. Injection of air at constant volume-flux reopens the initially collapsed channel via the propagation of an air finger. We observe various steadily propagating and transiently evolving fingers with different tip morphologies within the straight sections of channel. Moreover, the spatially varying channel width also leads to evolving finger shapes near the channel bifurcation. In some regimes, multiple reopening scenarios are recorded downstream of the channel branching, suggesting that the system is sensitive to small perturbations. This type of elasto-rigid channel has been used previously as an idealised model of pulmonary airway reopening during the first breath of a newborn, when an air bubble reopens the strongly collapsed, branched airway network filled with liquid secreted by the lung epithelium~\citep{Ducloue2017b, Juel2018}.
However, to our knowledge, this is the first attempt at studying experimentally the combined influence of channel bifurcation and elasticity in an airway reopening problem \citep{Heil15}. 

Attempts have been made to incorporate elastic elements into mostly rigid networks, and elasticity is known to affect finger propagation, for example, in the microfluidic Y-channel studied by \citet{Baroud2006}. In these rigid Y-channel experiments, asymmetric branching of the main finger was systematically observed when the identical daughter channels had open ends. However, when the daughter channels were each connected to a large elastic chamber, identical fingers propagated through both daughter channels for sufficiently small propagation speeds. The question of flow symmetry through different network pathways is fundamental to understanding flow in branching channels, and extends to the transport of bubbles, droplets, capsules and other kids of particles passing through a bifurcation in, e.g., microfluidics~\citep{mi11020201}. Another important question is how flow recovers after encountering the bifurcation, and what effect upstream bifurcations might have on those further downstream. For example, in laboratory models of blood flow, perturbations may propagate from bifurcation to bifurcation, because the multiphase flow does not have time to recover post-bifurcation before encountering a new fork in the network \citep{Merlo}. These fundamental questions call for an understanding of the role of elasticity in flow transport through a multi-generational branching network. 
 
So far, however, most theoretical and experimental works on airway reopening have focused on a single initially collapsed fluid-filled tube or compliant channel, reopened by the propagation of an injected air finger which redistributes the resident fluid. Early theoretical works \citep{HazelHeil2003, Halpern2005} systematically reported two branches of solution in terms of finger pressure $P$ and capillary number $Ca$, the dimensionless finger speed. The two branches met at a limit point associated with a minimum pressure, below which steady modes of reopening could not be computed. For higher values of pressure, two distinct behaviours were observed - pushing at low $Ca$, for which $P$ decreases with increased $Ca$, and peeling at high $Ca$, for which $P$ increases with increased $Ca$. Furthermore, the pushing branch was found to be linearly unstable in a two-dimensional (2-D) geometry consisting of a layer of liquid trapped between infinite, planar, elastic walls~\citep{Halpern2005}. Hence, stable reopening only took place via peeling solutions. However, despite revealing some of the fundamental dynamics of the reopening process, these early models were limited in scope: the 2-D model of \citet{Halpern2005} could not capture the in-plane finger shape because of its unbounded geometry, while the three-dimensional steady simulations of \citet{HazelHeil2003} imposed symmetry about the two longitudinal midplanes of the tube, thus eliminating a host of asymmetric solutions observed experimentally \citep{Heap2008}. 

The earliest experimental study of airway reopening reported a single peeling branch, for which a round-tipped finger of air injected at constant pressure reopened oil-filled polyethylene tubes collapsed into a ribbon-like configuration and held under axial tension \citep{Gaver1990}. However, it is now well established that tubes can be reopened by a variety of different propagation modes. In their experiments, \citet{Heap2008} studied tubes that were strongly collapsed into two-lobed cross-sections and revealed reopening characterised by round-, asymmetric-, double- and pointed-tipped finger shapes, with $P$ increasing with $Ca$ in all experiments. They also reported that at a critical value of $Ca$, a discontinuous drop in bubble pressure was accompanied by a transition from double- to pointed-tipped finger shapes, suggesting that the system had at least two discontinuous peeling branches of reopening. In an attempt to simplify the geometry, \citet{Ducloue2017b} explored reopening of a rectangular elasto-rigid channel of large aspect ratio (a Hele-Shaw channel) similar to the one studied in this paper, but without branching. This rendered the flow quasi-2-D, while maintaining much of the same physics of reopening as in collapsible tubes. For example, when the channel was initially collapsed so that a large proportion of the flexible sheet almost made contact with the rigid bottom boundary, \citet{Ducloue2017b} also observed a discontinuous drop in pressure at a critical $Ca$, which was accompanied by a change in the finger shape. 

\citet{Ducloue2017b} were the first to systematically study the reopening dynamics of an elasto-rigid channel by varying both initial collapse and $Ca$. The injection of air into an analogous liquid-filled rigid-walled channel of rectangular cross-section gives rise to the classical viscous fingering instability, which typically results in the steady propagation of a symmetric round-tipped finger \citep{SaffmanTaylor1958}. This displacement flow is governed by the ratio between viscous and surface tension forces measured by $Ca$ and it can also exhibit complex behaviours associated with, for example, geometric perturbations to the channel geometry \citep{Couder2000, Thompson2014, Franco-Gomez2016}.
In elasto-rigid channels, elastic forces work against the injected air finger in addition to viscous and surface tension forces, and this alters the two-phase flow fundamentally. Rather than displacing fluid, the propagating air finger tends to inflate the channel and redistributes fluid within a wedge ahead of its tip. This wedge is relatively long at lower levels of initial collapse and small $Ca$, and \citet{Ducloue2017b} found that symmetric fingers analogous to the Saffman--Taylor finger propagate in the channel. However, as the amount of fluid contained within the liquid wedge reduces with increasing initial collapse and/or $Ca$, fingers of different morphology can form as reported by \citet{Ducloue2017b} and explored further by \citet{Cuttle2020} and \citet{Fontana2022}. Longer liquid wedges are associated with dominant viscous and surface tension forces - this regime is referred to as `viscous', while shorter liquid wedges are linked to a dominant balance between elastic and surface tension forces - this regime is referred to as `elastic' \citep{Cuttle2020}. As the reopening transitions from viscous to elastic, fingers start penetrating the shallowest parts of the channel and peeling them apart much like in the classical problem of the peeling of a flexible strip \citep{McEwan1966, Lister2013}. Furthermore, their fronts become flatter and prone to interfacial instabilities \citep{Couder2000, PihlerPuzovic2012, Ducloue2017a}, leading to more exotic shapes of reopening fingers such as, e.g., feathered fingers, whose front perturbations are advected away from the tip along trajectories instantaneously normal to the curved interface \citep{Lajeunesse2000}. 

The growing complexity of the reopening modes of the elasto-rigid channel was captured in numerical simulations of \citet{Fontana2021, Fontana2022}, who developed a depth-averaged model of the system in the frame of reference of the
advancing air–liquid interface. At a fixed value of $Ca$ and increasing level of initial collapse, \citet{Fontana2021} found a complex solution structure with a wide range of stable and unstable, steady and time-periodic modes, many occurring at similar values of the driving pressure. 
A similar picture was observed at fixed levels of initial collapse and increasing $Ca$~\citep{Fontana2022}. Significance of unsteady reopening was also corroborated by experimental findings, e.g. that of \citet{Cuttle2020}, who found that at high levels of initial collapse steadily propagating fingers mapped onto two disconnected lines of bubble pressure increasing with $Ca$ (two peeling regimes), which were separated by a region of $Ca$ with no steady reopening; depending on its initial conditions, at a given value of the bubble pressure the system could assume either of the peeling behaviours. 

Given that unsteady reopening is common in a rectangular elasto-rigid channel with high level of initial collapse \citep{Fontana2021, Fontana2022}, it is also likely to be a feature of the bubble propagation in a network of elastic channels. However, in theoretical models of reopening of a network of occluded airways, only steady reopening is assumed to take place and only if the bubble pressure exceeds a critical threshold dependent on material parameters \citep{Stewart2015, Louf2020} in line with early studies of reopening by \citet{HazelHeil2003, Halpern2005}. Wall elasticity is also likely to be of fundamental importance in the related problem of liquid plug propagation in airways \citep{Yamaguchi2014}. However, this phenomenon is typically investigated using rigid branching channels \citep{Cassidy2001, Song2011, Baudoin2013}, in which the physics is described by relating the driving pressure head to the flow resistance caused by the viscous dissipation and capillarity. These can be written down accurately for each network generation if its geometry is known, but we are yet to build tractable models in which the network geometry can evolve in response to the changing fluid resistance as in elastic-walled channels.

In this paper, we explore experimentally the transfer of reopening modes between the channels upstream and downstream of the elasto-rigid Y-bifurcation at various levels of initial collapse and injection rates. Our experimental set-up and methods are introduced in \S~\ref{ExpMethods}. In \S~\ref{Section3}, we show that steady modes of finger propagation upstream of the Y-bifurcation transfer to different steady modes expected for suitably reduced flow rates downstream of the Y-bifurcation. However, in the limit of large initial channel collapse multiple reopening modes can coexist downstream of the Y-bifurcation, and we explore this regime further in \S~\ref{Section3.2}.
We also find that transient evolution of reopening fingers downstream of the Y-bifurcation typically persists over a significant distance before they recover a state of steady propagation. Hence, in \S~\ref{Section3.3} we study the unsteady finger propagation through the Y-bifurcation, and relate the time and length-scales for fingers to recover steady states in the daughter channels to the reopening dynamics. Our results suggest that simple transfer scenarios through the Y-bifurcation can only be assumed for sufficiently long channels and only in the regimes without finger multiplicity. Our conclusions are presented in \S~\ref{Conclusion}.

\section{Experimental methods}
\label{ExpMethods}

\begin{figure}
\centering
\includegraphics[width=12cm]{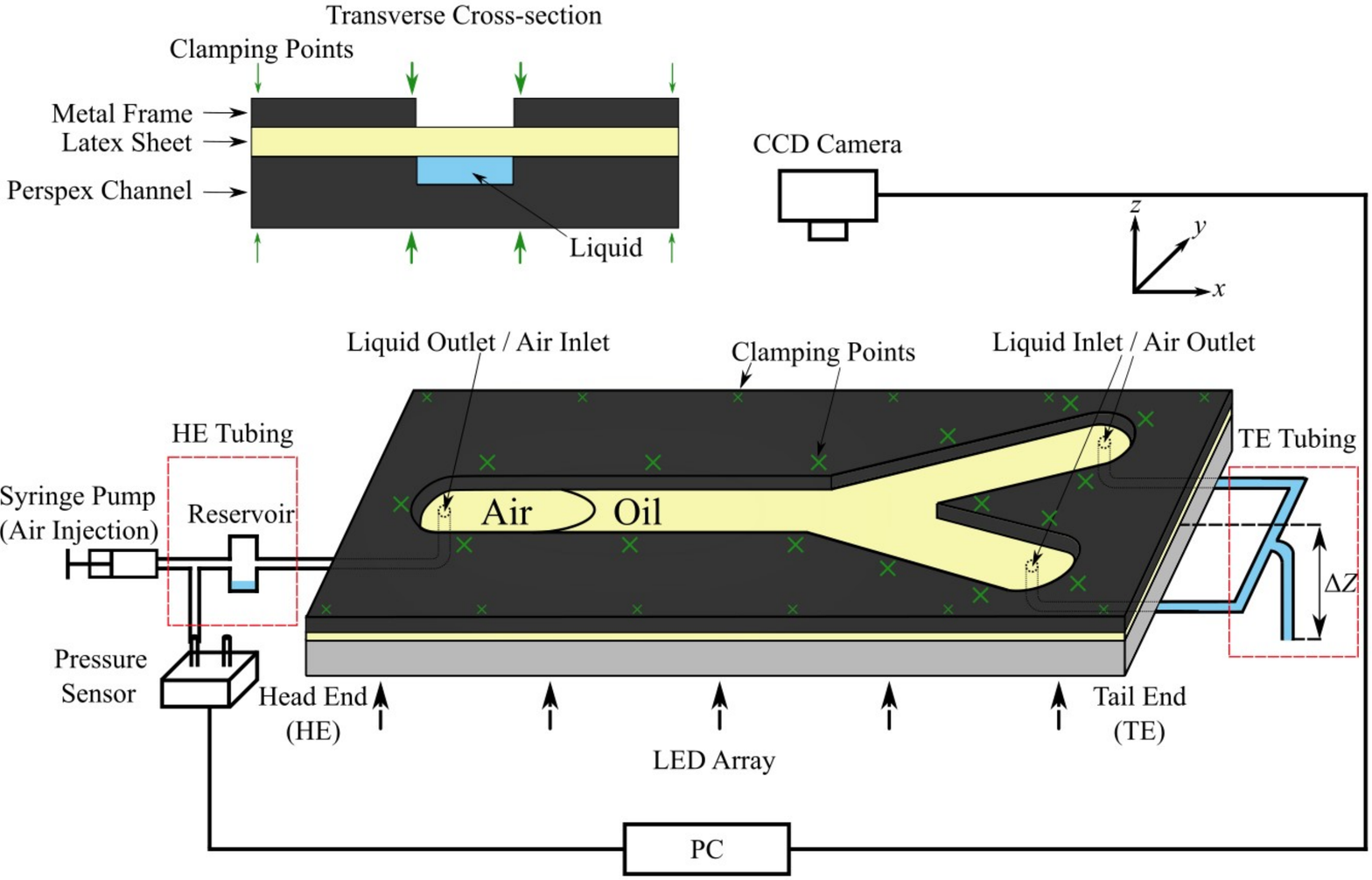}
\caption{Schematic diagram of the experimental set-up.}
\label{fig:Experimental Set-up}
\end{figure}

\begin{figure}
\centering
\includegraphics[width=0.8\linewidth]{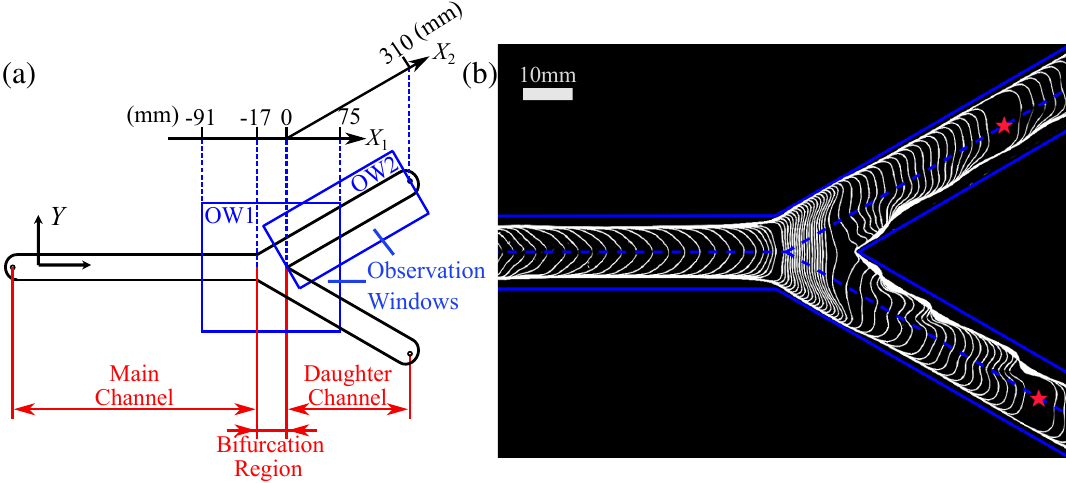}
\caption{(a) Schematic diagram of different regions in the Y-channel. $X_{1}$- and $X_{2}$-axis are aligned with the centrelines of the main and daughter channels, respectively, and the $Y$-axis is perpendicular to $X_{1}$, so that the apex of the bifurcation is at $X_{1}=X_{2}=Y=0$.
The main channel is of length 483 $\pm$ 0.02~mm and width 15 $\pm$ 0.02~mm, and the two daughter channels are of length 310 $\pm$ 0.02~mm and width 15 $\pm$ 0.02~mm each. They meet in the bifurcation region of length 17 $\pm$ 0.02~mm (i.e. the start of the region is at $X_{1}$ = -17~mm), in which the channel width varies linearly between 15 $\pm$ 0.02~mm and 34.64 $\pm$ 0.02~mm. (b) Typical spatiotemporal composite image of the bifurcation region, obtained for $A_i$ = 0.36 in the main channel and $Q$ = 50~ml/min. The time interval between sequential contours is 0.033~s. The blue solid and dashed lines denote the boundaries and the central lines of the main/daughter channels, respectively. In (b), one frame is lost during recording, so the position of the missing contour is marked with a red star.
}
\label{fig:OW}
\end{figure}

The experimental set-up, a Y-shaped elasto-rigid Hele-Shaw channel with a flexible upper boundary, is shown schematically in figure~\ref{fig:Experimental Set-up}. It consists of three regions: a main channel, two daughter channels and a bifurcation region, where the two daughter channels meet (see figure~\ref{fig:OW}(a)). The Y-shaped channel was milled into a transparent Perspex block by cutting a rectangular channel of length 500 $\pm$ 0.02~mm, width $W = 15 \pm$ 0.02~mm and depth $b = 0.5 \pm$ 0.01~mm, and two rectangular channels of length 310 $\pm$ 0.02~mm, with the same width $W$ and depth $b$ as the first channel. These were oriented at $\pm$ 30$^{\circ}$ to the centre line of the main channel. The Perspex block was covered by a rectangular latex membrane (Supatex) of thickness 0.33 $\pm$ 0.01~mm, Young’s modulus $E=$ 1.44 $\pm$ 0.05~MPa and Poisson’s ratio $\nu=$ 0.5 \citep{PihlerPuzovic2012}. Following the procedure developed by \cite{Ducloue2017b} and \citet{Cuttle2020}, the membrane was stretched uniformly in the direction transverse to the main channel and clamped using a metal frame and G-clamps to form the upper boundary of the Y-shaped channel. Uniform transverse pre-stress was achieved by hanging evenly-distributed weights of total mass 2.49 $\pm$ 0.01~kg onto the membrane along one of the long edges of the Perspex block before clamping. 

\begin{figure}
\centering
\includegraphics[width=0.8\linewidth]{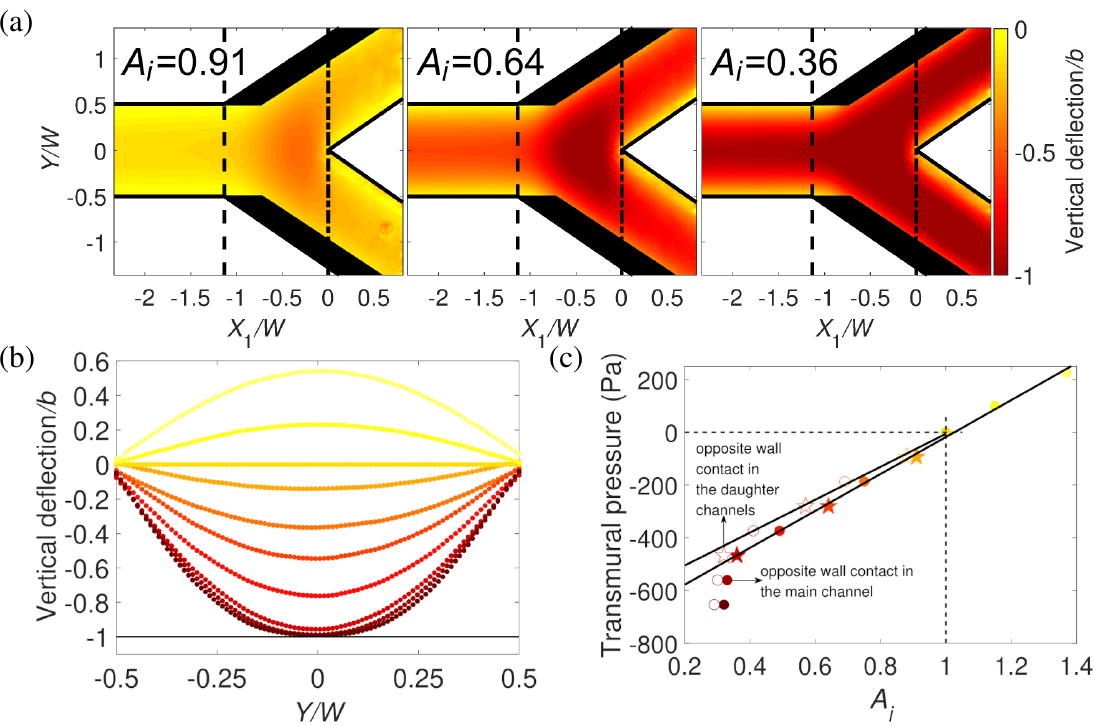}
\caption{(a) Membrane profiles for $A_{i}$ = 0.91, 0.64 and 0.36 in the main channel (left to right) before the start of air injection with colour-coding corresponding to the vertical deformation of the membrane scaled by $b$. Regions of the channel, where the vertical deformation could not be reconstructed, are black (for details, see Appendix \ref{appA}). Lines denote the boundaries of the channel (solid lines), the start of bifurcation region (vertical dashed line) and the apex of the bifurcation (vertical dash-dotted line). (b) Transverse profiles of the membrane at $X_{1}/W$ = -2 for $A_{i}$ = 1.37, 1.15, 1, 0.91, 0.75, 0.64, 0.49, 0.36, 0.33 and 0.32 (top to bottom) before the air injection starts. (c) Transmural pressure across the membrane as a function of the level of initial collapse $A_{i}$ in the main (filled markers) and daughter (empty markers) channels, with zero transmural pressure corresponding to no membrane deflection ($A_{i}$=1). Lines are linear fits to the experimental data above the points of the opposite wall contact. Star markers are used for the levels of collapse which correspond to the membrane  profiles in (a) and are referred to as low ($A_{i}$ = 0.91/0.88 in the main/daughter channels), moderate ($A_{i}$ = 0.64/0.57 in the main/daughter channels) and high ($A_{i}$ = 0.36/0.32 in the main/daughter channel) initial collapse in the paper, respectively. The colours in (b) and (c) are consistent.}
\label{fig:Ai}
\end{figure}

Prior to each experiment, the channel was filled with silicone oil (Basildon Chemical Company Ltd.) of viscosity $\mu=$ 0.019~Pa~s, density $\rho=$ 953~kg~m$^{-3}$ and surface tension $\sigma=$ 20.8~mN~m$^{-1}$ at temperature $T=$ 21 $\pm$ 1~$^{\circ}$C. The liquid/gas was supplied to/evacuated from the system using ports at the channel ends connected to tubing, and referred to as the head end (HE) and the tail end (TE) in figure~\ref{fig:Experimental Set-up}. During the filling procedure, some of the liquid drained from the main channel into the HE tubing, so a reservoir was added halfway along it to trap excess liquid. This avoided the obstruction of the air line with silicone oil. With the HE closed, the channel was collapsed by setting a hydrostatic pressure difference $P_{\mathrm{h}}$ between the channel and ambient laboratory pressure at the TE. The pressure difference caused the membrane to collapse symmetrically with respect to the vertical mid-plane of the channel and uniformly along most of its length (see figure \ref{fig:Ai}). 

The reopening experiments were performed by injecting air into the fluid-filled channel using a syringe pump (KDS 210) with a constant volumetric flux $Q$ set between 10~ml/min and 150~ml/min. The resulting propagation of a long air finger within the Y-shaped channel was captured in two observation windows with top-view CMOS cameras (Teledyne Dalsa CR-GM00-H1400, resolution 1400 $\times$ 1024~pixels) at 60 frames per second. To improve image quality, the channel was lit from below by a custom-built LED array. The image resolution was 0.117~mm/pixel in the first observation window, which covers the main channel, the bifurcation region and the upstream end of the daughter channels, and 0.224~mm/pixel in the second observation window, which covers just one of the daughter channels (see figure~\ref{fig:OW} (a)).
The images were processed in MATLAB to create spatiotemporal composite images, such as in figure~\ref{fig:OW}(b). During the longer recordings, occasional frames were lost due to technical limitations of the imaging hardware, and their positions are marked with red stars in the composite images. The spatiotemporal patterns enabled us to study finger evolution and extract, for example, the finger speed and tip position. The finger pressure $P$ (relative to atmospheric pressure) was measured with a pressure sensor (Honeywell 163PC01D36 $\pm$ 5 $^{\prime \prime}$ H$_{\rm 2}$O) inserted into the middle of the air supply line using a T-shaped tubing adapter, taking care to subtract the small pressure losses in the line. 

The time evolution of the membrane deformation during reopening experiments was reconstructed using a custom-built digital-image-correlation technique, described in detail in Appendix \ref{appA}. Examples of collapsed membrane profiles before the start of air injection are shown in figure~\ref{fig:Ai}(a). These results indicate that for each value of $P_{\mathrm{h}}$, the initial level of collapse differs in different regions of the Y-channel because of its varying geometry. For example, opposite wall contact, in which the thickness of the liquid film between the membrane and the channel becomes smaller than our experimental resolution, first occurs in the bifurcation region where the channel is the widest and thus has the smallest effective stiffness \citep{Audoly08}. 
The level of initial collapse is reported using the non-dimensional cross-sectional area $A_{i}$ = $A_{t}/A_{0}$ measured at $X_{1}/W$ = -2 in the main channel, where $A_{t}$ and $A_{0}$ denote the collapsed and undeformed transverse cross-sectional areas, respectively, so that $A_{t}/A_{0}=1$ corresponds to a channel with a rectangular cross-section. % and is the neutral point of the main channel.

Examples of membrane profiles in the cross-section of the main channel, which are used for estimating $A_{i}$, are shown in figure~\ref{fig:Ai}(b) for a range of transmural pressures (i.e. the difference between internal and external pressure). The shape of the channel is symmetric with respect to its centreline where the deflection is maximum.  The channel is inflated under positive transmural pressure ($A_{t}/A_{0}>1$), and collapsed under negative transmural pressure  ($A_{t}/A_{0}<1$). The same data is presented differently in figure~\ref{fig:Ai}(c), where transmural pressure is shown as a function of initial level of collapse for the main channel.  Figure~\ref{fig:Ai}(c) also includes data for the daughter channels taken at $X_2/W=0.67$. It demonstrates that for a fixed value of transmural pressure, the daughter channels are more collapsed (i.e. have a smaller $A_i$) than the main channel. This is because the pre-stretch is applied in the direction normal to the main channel and thus obliquely to the daughter channels, which means that the daughter channels have a smaller effective stiffness. Based on estimates using the finite number of points in figure~\ref{fig:Ai}(c), the first opposite wall contact in the main channel occurs at $A_{i}$ = 0.33, whereas it is already seen in the daughter channels at $A_{i} = 0.36$. In the vicinity of those points, small changes in $A_{i}$ result in large pressure variations in the main and daughter channels, respectively. However, for the data above the points of opposite wall contact in figure~\ref{fig:Ai}(c), the transmural pressure increases linearly with $A_i$. In this paper, we mostly report experiments at $A_{i}$ = 0.91, 0.64 and 0.36 in the main channel, and refer to these levels of initial collapse as low, moderate and high, respectively. The corresponding $A_{i}$ in the daughter channels were 0.88, 0.57 and 0.32, respectively, and we stress the difference between the initial cross-sectional areas in different parts of the Y-channel where required.

\section{Results}
\label{Results}

\begin{figure}
\centering
\includegraphics[width=0.6\linewidth]{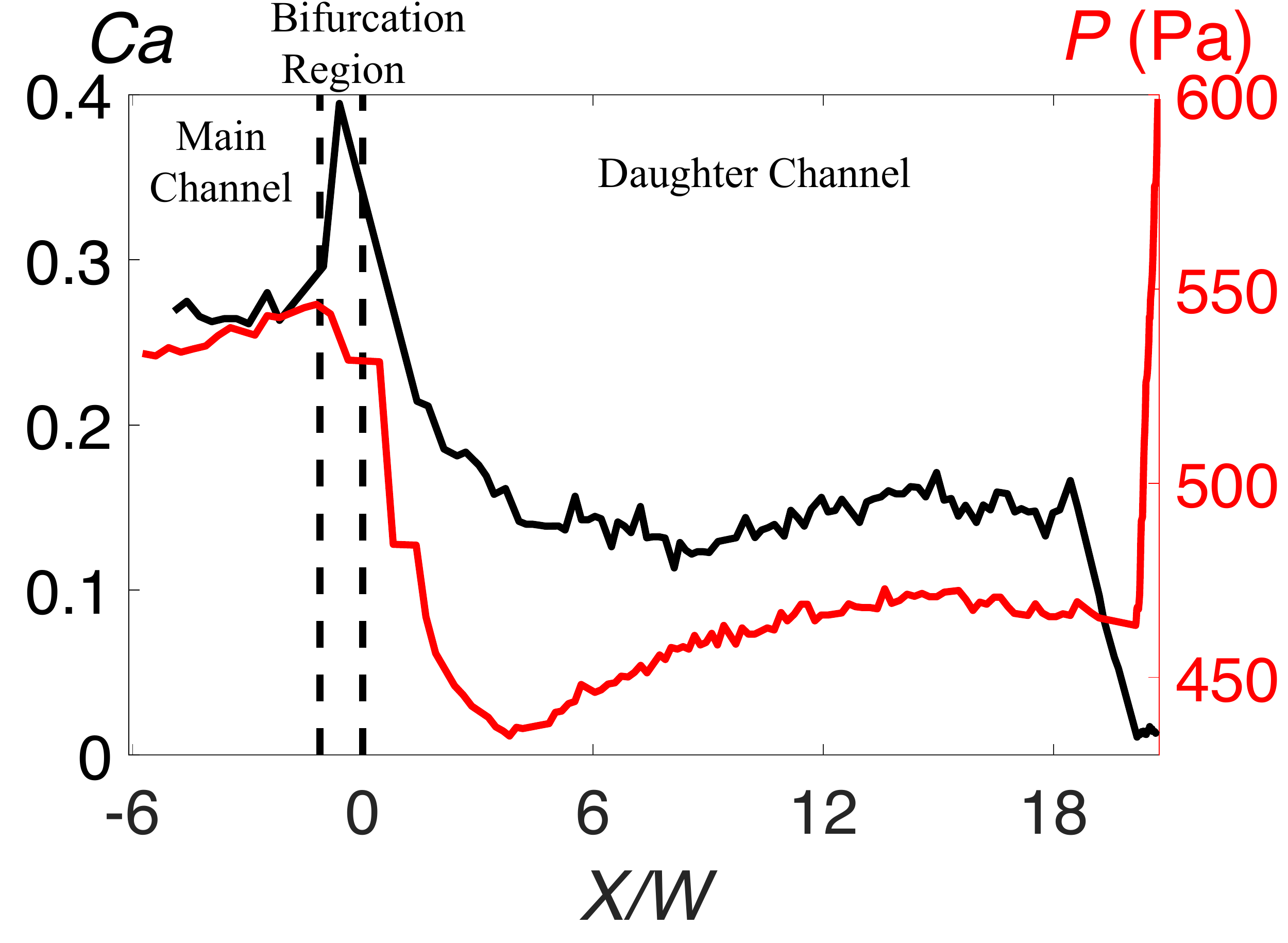}
\caption{(a) $Ca$ (black line) and (c) $P$ (red line) as a function of scaled finger tip position $X/W$ ($X=X_1\cup X_2$) for $A_i=0.91/0.88$ in the main/daughter channels and $Q =$ 150 ml/min. Vertical dashed lines delineate different regions in the Y-channel.}
\label{fig:CaPX_overall}
\end{figure}

We performed reopening experiments in our Y-shaped channel under low, moderate and high levels of initial collapse (see \S \ref{ExpMethods}), for a wide range of constant volumetric air injection rates $Q$. Figure~\ref{fig:CaPX_overall} shows typical traces of the dimensionless finger speed $Ca$ (black line) and bubble pressure $P$ (red line) as a function of the position of the finger tip $X$ normalised by the channel width $W$. This finger tip coordinate is aligned in turn with the main and daughter channels as the air finger propagates through the Y-shaped channel. The pressure is approximately constant to within the experimental resolution of 5~Pa in the main channel. This indicates a steady mode of finger propagation, which exhibits an approximately constant $Ca$. The varying channel geometry around the Y-bifurcation leads to non-monotonic variations of $Ca$ and $P$ as the finger tip approaches and propagates through the bifurcation region. The main finger splits at the Y-bifurcation into two propagating daughter fingers with the same pressure $P$ and individually measured values of $Ca$. As they propagate through the daughter channels, their typical evolution is towards a steadily propagating finger. However, in the experiment shown in figure~\ref{fig:CaPX_overall}, both $P$ and $Ca$ exhibit measurable variations in the first half of the daughter channel, only reaching constant values for $X_2/W\approx 12$. 
In the daughter channel, the finger propagates steadily with a value of $Ca$ that is approximately half of that in the main channel. This is because the flow rate divides equally between the identical daughter channels, which have very similar cross-sections to the main channel. In addition, the relationship between flow rate and $Ca$ is approximately linear \citep{Fontana2022}. The steady-state pressure is accordingly reduced in the daughter channels, compared to that of the main channel.

Our experiments also showed a small, systematic bias that promoted finger propagation in one daughter channel ahead of the other upon exit from the Y-bifurcation. However, once steady propagation was recovered, the relative position of the two finger tips did not change. Hence, the leading daughter finger reached the end of its channel before the trailing finger. Fluid ahead of the leading finger was displaced into the rigid outlet tubing, increasing resistance to air propagation. This led to a sharp increase of $P$ and the abrupt reduction of the velocity of the trailing finger (at $X_2/W \approx 18.4$) seen in figure~\ref{fig:CaPX_overall}. The distance from this point to the end of the daughter channel gives the relative distance between the tips of the leading and trailing daughter fingers in steady propagation. Although this distance varied with flow parameters, it remained systematically smaller than two channel widths and did not measurably influence the dynamics. Thus, we do not mention this bias henceforth. 

We begin in \S \ref{Section3} by examining how steady modes of finger propagation transfer from the main channel to the daughter channels. We discuss the multiple modes of propagation encountered as a function of $A_i$ and $Ca$ by comparison with previous studies in straight elasto-rigid Hele-Shaw channels \citep{Ducloue2017b, Cuttle2020, Fontana2021, Fontana2022}. We show in \S \ref{Section3.2} that for high collapse, reopening modes in the main channel can transfer to multiple outcomes in the daughter channels. We then proceed in \S \ref{Section3.3} to characterise the mechanics of unsteady reopening in the bifurcation region (\S \ref{Section3.3.1}) and its influence on the recovery of steady finger propagation in the daughter channels (\S \ref{Section3.3.2}).

\subsection{Steadily propagating fingers in the main and daughter channels}
\label{Section3}

\begin{figure}
\centering
\includegraphics[width=\linewidth]{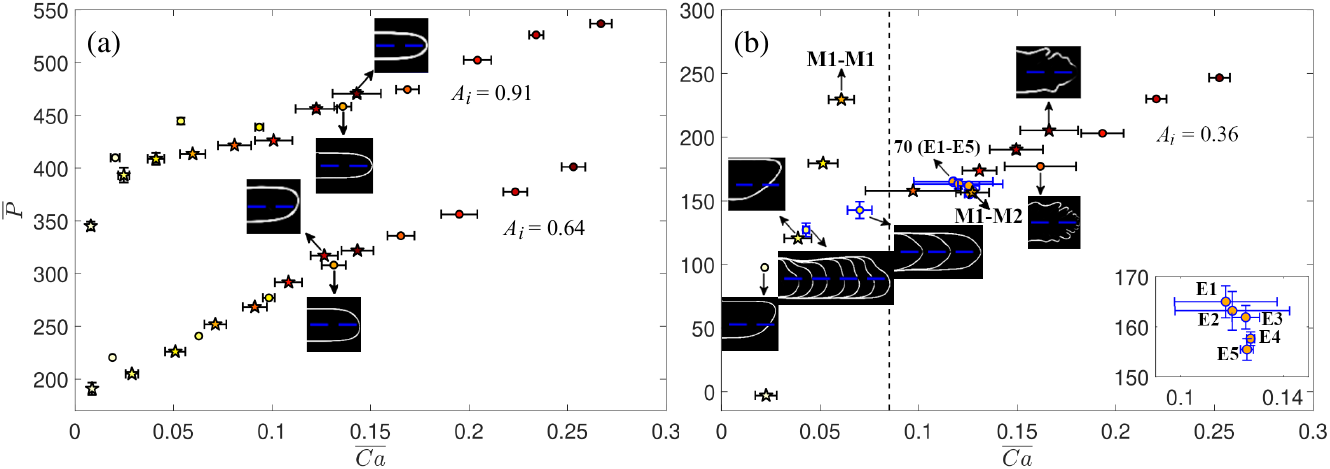}
\caption{Averaged finger pressure $\overline{P}$ as a function of averaged capillary number
  $\overline{Ca}$ for different levels of initial collapse: (a) $A_i = 0.91/0.88$ and $A_i = 0.64/0.57$ and (b) $A_i = 0.36/0.32$ in the main channel (circles) and the daughter channels (stars), respectively. 
The data for the main and the daughter channels obtained in the same experiment are shown using the same colours. 
Markers with blue boundaries in (b) correspond to transiently evolving fingers with a non-constant pressure trace throughout the main channel. 
The insets show typical fingering patterns (either steadily propagating or, when there is more than one contour, transiently evolving fingers) in different regions of the parameter space. The vertical line in (b), intended to guide the eye, separates the region of lower $\overline{Ca}$, in which the difference between the data in the main and the daughter channels is the greatest.
Points E1-E5 and M1-M1 and M1-M2 in (b), all obtained for $Q=70$~ml/min, are discussed in section~\ref{Section3.2}. The inset in the corner of (b) shows the enlarged region of $\overline{Ca} - \overline{P}$ containing points E1-E5.}
\label{fig:CaP_main}
\end{figure}

We found that fingers propagate steadily through the main channel for most of the parameters investigated. Their pressure, averaged over the domain $-3.5 < X_1/W < -2.5$, is shown with circles in figure~\ref{fig:CaP_main} as a function of the average capillary number $\overline{Ca}$, for low, medium and high initial collapse of the channel. 
The error bars on $\overline{P}$, which indicate standard deviation over the visualisation domain, are always within $\pm 7$~Pa and thus typically hidden by the markers. The largest pressure fluctuations were measured for fingers which failed to settle to a state of steady propagation and thus, evolved continually over the visualisation window. These experiments are shown with blue markers in figure~\ref{fig:CaP_main}. In contrast, the standard deviation of $\overline{Ca}$ is not a good indicator of steady finger propagation. The corresponding error bars tend to increase as $Ca$ or the level of initial collapse increases, because the liquid wedge ahead of the finger tip which sets the finger velocity becomes thinner and thus, the finger becomes increasingly sensitive to channel imperfections \citep{Ducloue2017a, Cuttle2020}. 

Stars indicate fingers propagating in the daughter channels, which are shifted to approximately half the value of $\overline{Ca}$ in the main channel. The colour of the markers is used to identify fingers in the main and daughter channels which originated from the same experiment. The length of the region where fingers propagate steadily depends on the length of transients but averages of $\overline P$ and $\overline Ca$ are taken after the finger reaches steady propagation (see section~\ref{Section3.3}).

Results shown in figure \ref{fig:CaP_main} are consistent with previous studies in non-branching channels, with small quantitative differences originating in variations of the constitutive relationships between pressure and cross-sectional area of the channels \citep{Cuttle2020,Fontana2021,Fontana2022}. In all experiments, the finger pressure increases with $\overline{Ca}$, which indicates reopening via the propagation of peeling fingers \citep{Gaver1996}. In figure \ref{fig:CaP_main}(a), which shows results for low and medium levels of initial collapse, circles and stars follow similar trends indicating that the reopening dynamics are insensitive to the difference of up to 11\% between the initial level of collapse in the main and daughter channels (see \S \ref{ExpMethods}). This is confirmed by the inset images of matching fingers from the main and daughter channels at the same values of $\overline{Ca}$. For low and medium levels of initial collapse in our experiments, the fingers are round-tipped. They are associated with a relatively long liquid wedge ahead of them, so that the dominant force balance is between viscous and surface tension forces. At $A_{i}$ = 0.91/0.88 in the main/daughter channel, the fingers are symmetric about the centreline of the channel. When $A_{i}$ = 0.64/0.57 in the main/daughter channel, the fingers become increasingly asymmetric about the centreline of the channel with increasing $\overline{Ca}$. 
As explained by \cite{Ducloue2017a}, the finger asymmetry at moderate levels of collapse stems from more significant variations of the liquid thickness within the transverse cross-section of the channel: the maximum collapse of the membrane occurs along the centreline of the channel, with deflection reducing to zero near its boundaries where the sheet is clamped~(see figure~\ref{fig:Ai}(b)). This promotes reopening with fingers propagating closer to one of the channel side walls where the viscous resistance is smaller.
 
We observed more complex dynamics for high initial collapse (figure~\ref{fig:CaP_main}(b)) consistent with previous findings \citep{Cuttle2020, Fontana2021,Fontana2022}.
In the main channel ($A_{i} = 0.36$), a succession of different modes of finger propagation was observed for increasing $\overline{Ca}$, including asymmetric, asymmetric with intermittent tip-perturbations, pointed fingers and feathered patterns, where small-amplitude viscous fingering occurs at the bubble tip. The appearance of feathered patterns has been associated with a switch from viscous forces to elastic forces balancing surfaces tension forces in the reopening process. Figure~\ref{fig:CaP_main}(b) also shows that the dynamics at high initial collapse (i.e., close to the point of the opposite wall contact) is very sensitive to small changes of the initial cross-sectional area between the main and daughter channels. In the main channel, all the reopening fingers exhibit unsteady behaviour in the vicinity of the vertical dashed line, indicating a region of complex dynamics. For $\overline{Ca}\lesssim 0.09$ (left of the vertical dashed line in figure~\ref{fig:CaP_main}(b)), the finger pressure in the daughter channel increases steeply with increasing $\overline{Ca}$ and does not map to the peeling curve measured for the main channel. Accordingly, the inset pictures captured in the main and daughter channels at approximately the same value of $\overline{Ca}$ do not match like in figure \ref{fig:CaP_main}(a). 
In contrast, for $\overline{Ca} \gtrsim 0.09$ (right of the vertical dashed line in figure~\ref{fig:CaP_main}(b)), $\overline{P}$ varies similarly with $\overline{Ca}$ in both main and daughter channels.

Hence, figure~\ref{fig:CaP_main}(b) indicates that for the daughter channels, there are two disconnected branches of steadily propagating fingers on either side of the vertical dashed line. This is significant in that it leads to a region of bistability in the range $156 < \overline{P} < 231$~Pa, where fingers with similar values of $\overline{P}$ can propagate in the daughter channels with different average dimensionless speeds $\overline{Ca}$.

The importance of this region of bistability for the reopening of the Y-bifurcation is best illustrated by repeated experiments conducted at the flow rate $Q =70$~ml/min. The corresponding data points obtained in the main channel are marked as E1-E5 in figure~\ref{fig:CaP_main}(b). In these experiments, similar values of $\overline{P}$ and $\overline{Ca}$ were recorded to within experimental resolution. However, propagation of the fingers into the daughter channel ($A_{i}$ = 0.32) led to two distinct peeling behaviours in consecutive experiments quantified by pressures that differed by 50\%. These two pressure values labelled M1-M1 and M1-M2 in figure \ref{fig:CaP_main}(b) are also associated with distinct values of the capillary number which lie on both sides of the dashed line. We explore the sensitivity of the reopening modes of the Y-channel in further detail in the next section (\S \ref{Section3.2}.)

\subsection{Multiple modes of finger propagation for the same injection rate}
\label{Section3.2}

\begin{figure}
 \centering
 \includegraphics[width=\linewidth]{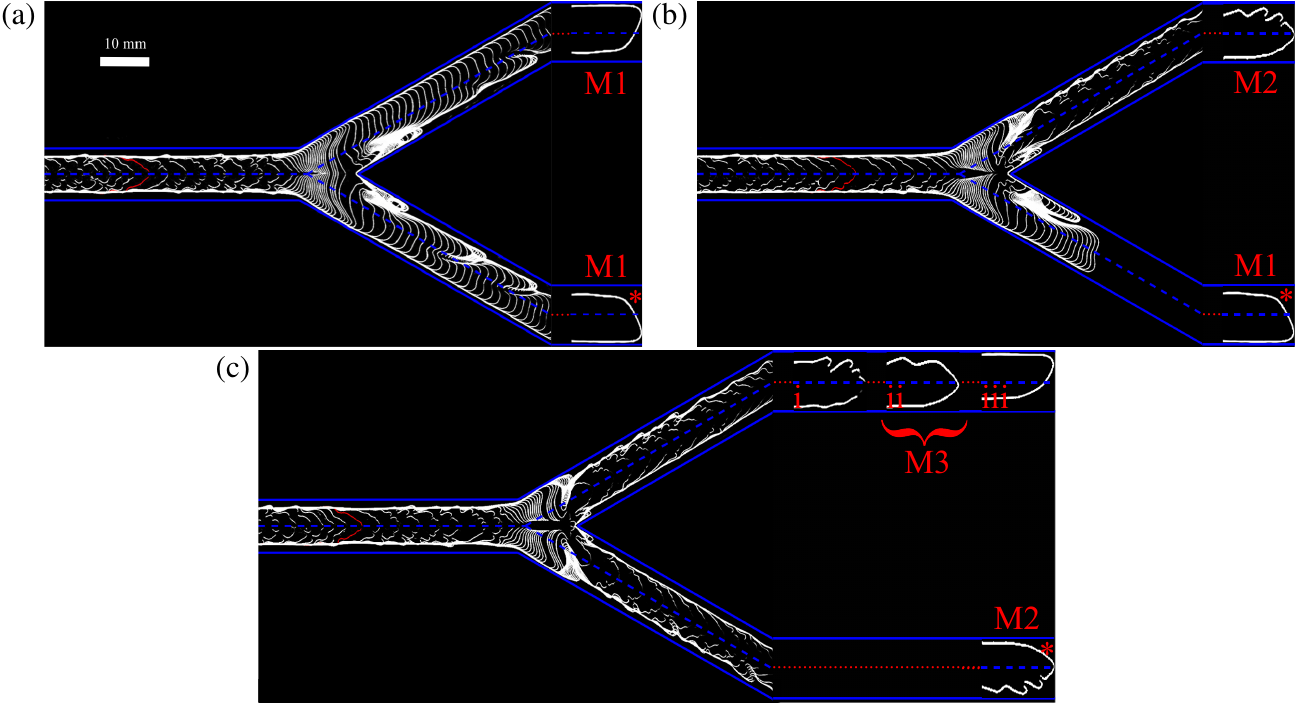}
\caption{Superposition of sequential contours with time interval of 0.033 s during reopening experiments, marked as (a) E1, (b) E2 and (c) E3 in figure~\ref{fig:CaP_main}, at $Q$ = 70~ml/min and $A_{i}$ = 0.36 in the main channel. Insets show typical instantaneous finger shapes, marked as M1, M2 and M3, during reopening of the daughter channels, initially collapsed to $A_{i}$ = 0.32. The shapes with asterisks were not recorded directly; instead, snapshots from the other daughter channel are used as representations of the shapes observed.}
\label{fig:MultipleModes}
\end{figure} 

Figure~\ref{fig:MultipleModes} shows the different modes of reopening observed in twenty experiments conducted at a flow rate $Q =70$~ml/min and high initial collapse ($A_i=0.36$). 
The fingers in the main channel are feathered, but variability in the observed patterns indicates the transient evolution of the reopening fingers over the visualisation region.
These fingers even briefly resemble the pointed symmetric fingers observed at lower values of $\overline{Ca}$ (compare the red contours in figure~\ref{fig:MultipleModes} to finger shapes in figure~\ref{fig:CaP_main}(b) at $\overline{Ca}\approx 0.075$).
The insets on the right hand-side of figure~\ref{fig:MultipleModes} show three types of fingers, labelled M1, M2 and M3, that were observed in the daughter channels in all of the twenty experiments following recovery downstream of the Y-bifurcation. 
M1 and M2 correspond to asymmetric fingers and feathered fingering patterns, respectively.
M3 fingers continually evolved throughout the daughter channel from (i) a feathered finger to (ii) a pointed finger and finally to (iii) an asymmetric finger. Remarkably, the two daughter channels could be either reopened by the same mode, e.g., M1 in figure~\ref{fig:MultipleModes}(a), or different modes of finger propagation, e.g. M1 and M2 in figure~\ref{fig:MultipleModes}(b) or M2 and M3 in figure~\ref{fig:MultipleModes}(c). We label these experiments according to the fingering modes observed in the daughter channels, e.g. M1-M2 corresponds to the reopening in figure~\ref{fig:MultipleModes}(b).   

\begin{figure}
 \centering
 \includegraphics[width=\linewidth]{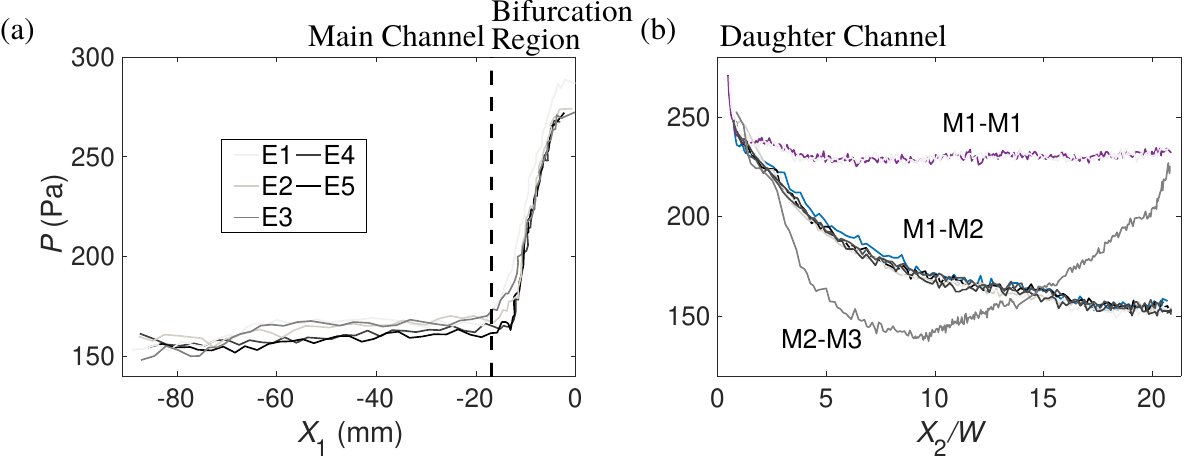}
\caption{Finger pressure $P$ during the channel reopening as a function of scaled finger tip position (a) in the main channel $X_{1}/W$ and (b) in the daughter channel $X_{2}/W$, when $A_{i}$ = 0.36/0.32 in the main/daughter channel and $Q$ = 70~ml/min. The vertical dashed line in (a) denotes the start of the bifurcation region.}
\label{fig:MultipleModes_P}
\end{figure}

The pressure data in the main channel and bifurcation region, shown in figure~\ref{fig:MultipleModes_P}(a) for five of the experiments at $Q=70$~ml/min, indicates small variability between experiments on the order of the experimental resolution. All pressure traces take approximately the same average value of 160$\pm 7$~Pa upstream of the Y-bifurcation and increase steeply within the bifurcation region as previously shown in figure \ref{fig:CaPX_overall}. In fact, some of the pressure traces are literally superposed, like E4 and E5. However, these small differences between experimental runs are sufficient to yield qualitatively different pressure traces downstream of the bifurcation region.

In the daughter channels, the reopening experiments introduced in figure \ref{fig:MultipleModes} are each associated with a qualitatively different pressure trace. Pressure traces from 10 experiments are shown in figure \ref{fig:MultipleModes_P}(b), with the remaining data not shown to avoid overloading the figure.
Strikingly, a steady state of constant pressure was reached almost immediately for experiments labelled M1-M1, whereas for the experiments labelled M1-M2, it was only reached at the end of the daughter channel with a considerably lower value. Although the final pressures for M1-M1 and M2-M3 were similar, the M2-M3 pressure varied non-monotonically, consistent with the evolution of the M3 finger shapes shown in figure~\ref{fig:MultipleModes}. (Because of this transient reopening throughout the daughter channel, the experimental data for M2-M3 is not indicated in figure~\ref{fig:CaP_main}(b)).

These experiments clearly show that the modes of reopening of the daughter channels cannot always be predicted from knowledge of the finger patterns and pressure traces in the main channel. Beyond the multiplicity of long-term reopening scenarios, the multiple routes to steady-state propagation are also associated with a variety of characteristic time- and length-scales for the decay of transients. Although steady propagation was usually reached within the finite length of the daughter channels, the dynamics of the propagating finger would sometimes also exhibit long-lived transients during which the finger appeared to successively explore different modes of propagation known to exist in this system \citep{Fontana2021}. 

\subsection{Unsteady finger propagation in the Y-channel}
\label{Section3.3}

We have found that the recovery distance of a steadily propagating finger post-bifurcation can range from very short to longer than the daughter channel. This suggests that transient finger evolution during recovery may be more important to the reopening of an elastic network than steady states of propagation that may only be reached in sufficiently long channels. 
In this section, we demonstrate that finger propagation through the bifurcation region and the steady-state recovery in the daughter channels are determined primarily by the volume of the liquid wedge ahead of the finger and thus, whether reopening is dominated by viscous or elastic resistance to finger propagation.

\subsubsection{Reopening of the bifurcation region}
\label{Section3.3.1}

\begin{figure}
  \vspace{4mm}
	\centering
	\includegraphics[width=0.8\linewidth]{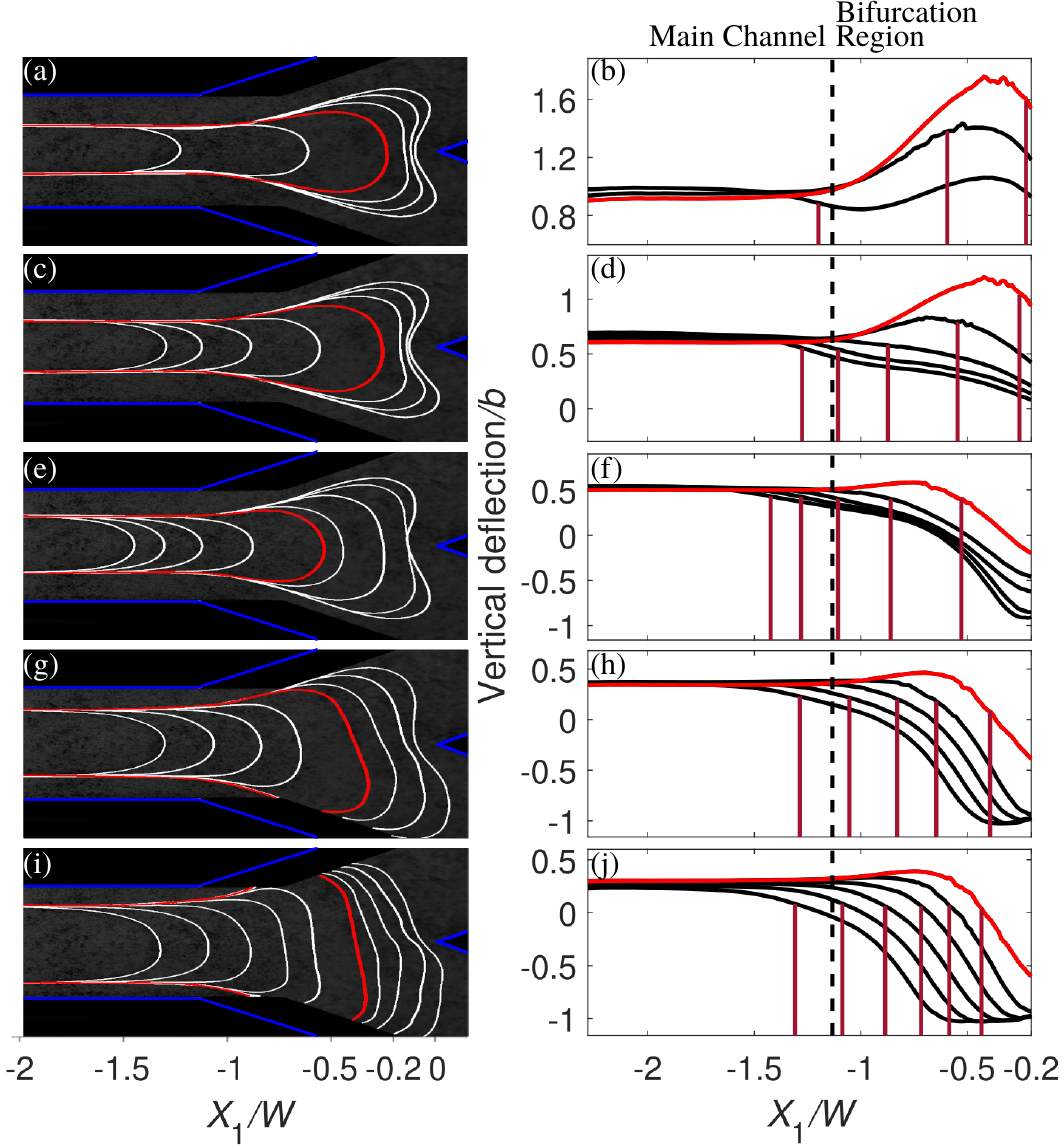}
	\caption[4]{Finger shapes (left column), instantaneous membrane deformations in the longitudinal mid-plane of the main channel (black lines in the right column) and the corresponding positions of the finger tip (red vertical lines in the right column) for (a, b) $A_{i}$ = 0.91, (c, d) $A_{i}$ = 0.75, (e, f) $A_{i}$ = 0.64, (g, h) $A_{i}$ = 0.49 and  (i, j) $A_{i}$ = 0.36 and $Q$ = 10~mL/min (all levels of initial collapse cited based on the main channel $A_i$). Note that there are more finger contours than corresponding membrane deformations due to limited resolution of the membrane deformation reconstruction method near the apex of the bifurcation ($-0.2<X_1/W<0$), see Appendix \ref{appA}; the last contours and membrane deformations that correspond to each other are shown in red. Time intervals between sequential contours/membrane deformations are (a,~b)~0.08~s, (c,~d)~0.04~s, (e,~f)~0.08~s and then~0.3~s beyond the red contour, (g,~h)~0.24~s and then~0.43~s beyond the red contour, and (i,~j)~0.36~s and then~0.4~s beyond the red contour,
 respectively. The blue lines in the left column mark boundaries of the Y-channel and the dashed line in the right column marks the start of the bifurcation region.}
	\label{fig:MembraneDeformationReopening}
\end{figure}
      
We begin by examining the reopening of the bifurcation region where the channel widens ahead of the apex of the Y-bifurcation. The height of the collapsed channel is lower in this region compared with the main channel, as previously shown in figure~\ref{fig:Ai}(a). We focus on experiments performed at a modest value of the flow rate $Q$ = 10~ml/min so that both main and daughter channels can reopen steadily for all levels of initial collapse. 

Figure~\ref{fig:MembraneDeformationReopening} shows sequences of finger outlines in top-view (left column) and the corresponding sheet profiles along the longitudinal vertical mid-plane of the main channel (black lines, right column) for five levels of initial collapse, $0.36 \le A_i \le 0.91$ in the main channel. In the side-view visualisation, the position of the propagating finger tip is indicated by vertical red lines.
For $0.64 \le A_{i} \le 0.91$, the moderate initial collapse means that the volume of fluid displaced in front of the finger as it propagates into the bifurcation region is sufficient to inflate this domain significantly beyond the level of inflation in the main channel (figure~\ref{fig:MembraneDeformationReopening}(b,d,f)). This excess inflation is enhanced by the slowing of the finger as it approaches the apex of the bifurcation. Thus, the fingers retain the round tip characteristic of viscous reopening through the bifurcation region, until the interface flattens to accommodate tip-splitting caused by the forking of the channel (figure~\ref{fig:MembraneDeformationReopening}(a,c,e)).  However, as $A_i$ is reduced, membrane profiles that correspond to fingers still propagating in the main channel get steeper, indicating a reduction in the length of the liquid wedge. When the level of initial collapse is reduced to $A_i=0.49$ and then further to $0.36$ (figure~\ref{fig:MembraneDeformationReopening}(g--j)), there is initial opposite wall contact between the top elastic sheet and the bottom boundary in the bifurcation region. This further reduces the volume of fluid during the reopening experiments, resulting in a minimal liquid wedge ahead of the finger which can no longer promote the inflation of the channel in the bifurcation region. As the finger advances into the bifurcation region with the elastic sheet in this limiting configuration (figure~\ref{fig:MembraneDeformationReopening}(h,j)) the finger tip flattens and its speed decreases slightly to accommodate the widening of the finger (figure~\ref{fig:MembraneDeformationReopening}(g,i)).

\begin{figure}
	\centering
	\includegraphics[width=\linewidth]{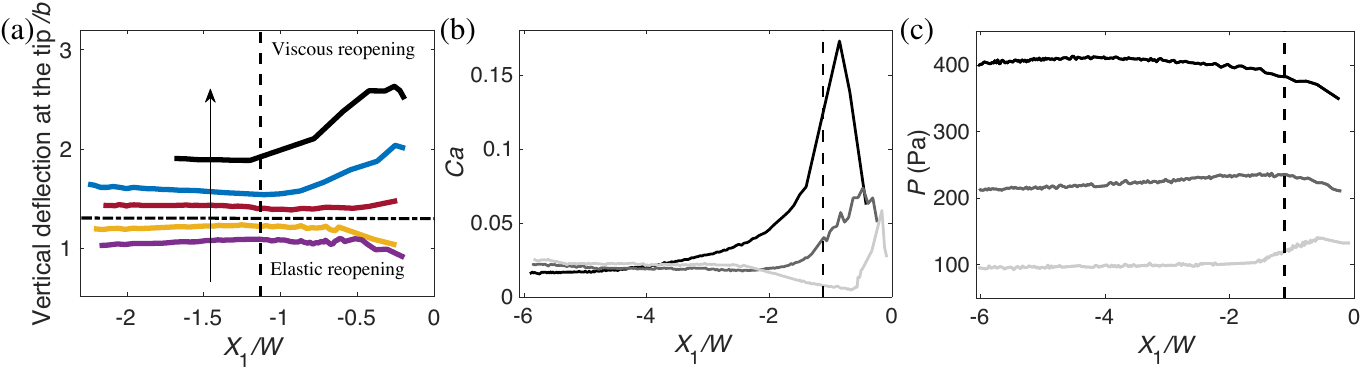}
	\caption[9]{(a) The membrane deflection at the finger tip, normalised by the channel depth $d$, as a function of scaled finger tip position $X_{1}/W$ for $A_{i}$ = 0.36, 0.49,
		0.64, 0.75, 0.91 in the main channel (increasing in the direction of the arrow) and $Q$ = 10~ml/min. The horizontal dashed line is added to stress the change in the behaviour of $h$. (b) $Ca$ and (c) $P$ as a function of scaled finger tip position $X_1/W$ for different $A_i$ and fixed $Q =$ 10 ml/min. The black, grey and light grey lines correspond to the experimental data for $A_i =$ 0.91, 0.64 and 0.36 in the main channel, respectively. The vertical dashed lines mark the start of the bifurcation region.}
	\label{fig:ViscousElastic}
      \end{figure}
      
Figure~\ref{fig:ViscousElastic} summarises the transition from viscous to elastic reopening depicted in figure~\ref{fig:MembraneDeformationReopening}. The sheet deflection at the finger tip shown in Figure~\ref{fig:ViscousElastic}(a) increases within the bifurcation region indicating enhanced inflation relative to the main channel for $0.64 \le A_{i} \le 0.91$. In contrast, a marginal decrease indicating deflation relative to the main channel occurs  for $0.36 \le A_{i} \le 0.49$.

The plots of $Ca$ and $P$ as a function of scaled finger tip position $X_1/W$ for high, medium and low collapse (Figures~\ref{fig:ViscousElastic}(b,c)) indicate that the geometry of the bifurcation region affects the finger while it is still propagating in the main channel. The distance at which departure from steady-propagation occurs increases with decreasing level of initial collapse because of the lengthening wedge of recirculated liquid ahead the finger. Pressure and capillary number traces also reflect the qualitative difference in the reopening of the bifurcation region between modest (black and grey curves) and high initial collapse (light grey curve). 
For low to moderate initial collapse, $Ca$ increases and $P$ decreases as the finger approaches the bifurcation region. This is because resistance to propagation is reduced in the widening channel which lowers viscous and elastic forces. Air compression is negligible for our parameters, so the finger propagation is driven by a constant air mass-flux. Therefore, once the finger enters the bifurcation region and begins to widen, $Ca$ decreases to satisfy air mass conservation. The wider channel means that lower pressure is required to inflate it and thus, pressure continues to decrease. In contrast, for $A_i=0.36$ in the main channel, the cross-sectional area of the channel does not change significantly as the finger enters the bifurcation region because the channel is highly collapsed. Once opposite wall contact occurs, a considerable increase in the transmural pressure is required to alter the channel cross section (see figure~\ref{fig:Ai}(c)). Thus, the finger quickly widens upon entering the bifurcation region but its thickness hardly changes which results in a weakly decreasing $Ca$ under the constant air mass-flux constraint. The finger eventually inverts the concave elastic sheet in a snapping-like manner at a threshold pressure. Once the sheet has inflated, $P$ and $Ca$ follow the trends observed for lower initial collapse.

The measurements presented in this section also suggest that reopening in different parts of the Y-channel can be dominated by different forces within the same experiment. For example, all fingers shown in figure~\ref{fig:MembraneDeformationReopening} have rounded tips in the main channel, which is indicative of viscous reopening. Yet, their propagation through the bifurcation region at larger levels of initial collapse (figure~\ref{fig:MembraneDeformationReopening}(g-j)) is typical of elastic reopening. Different reopening regimes within the same experiment are a consequence of the changing geometry, which leads to altered constitutive relationships between pressure and cross-sectional area of the channels.  

\subsubsection{Steady-state recovery in the daughter channels}
\label{Section3.3.2}

\begin{figure}
	\centering
	\includegraphics[width=\linewidth]{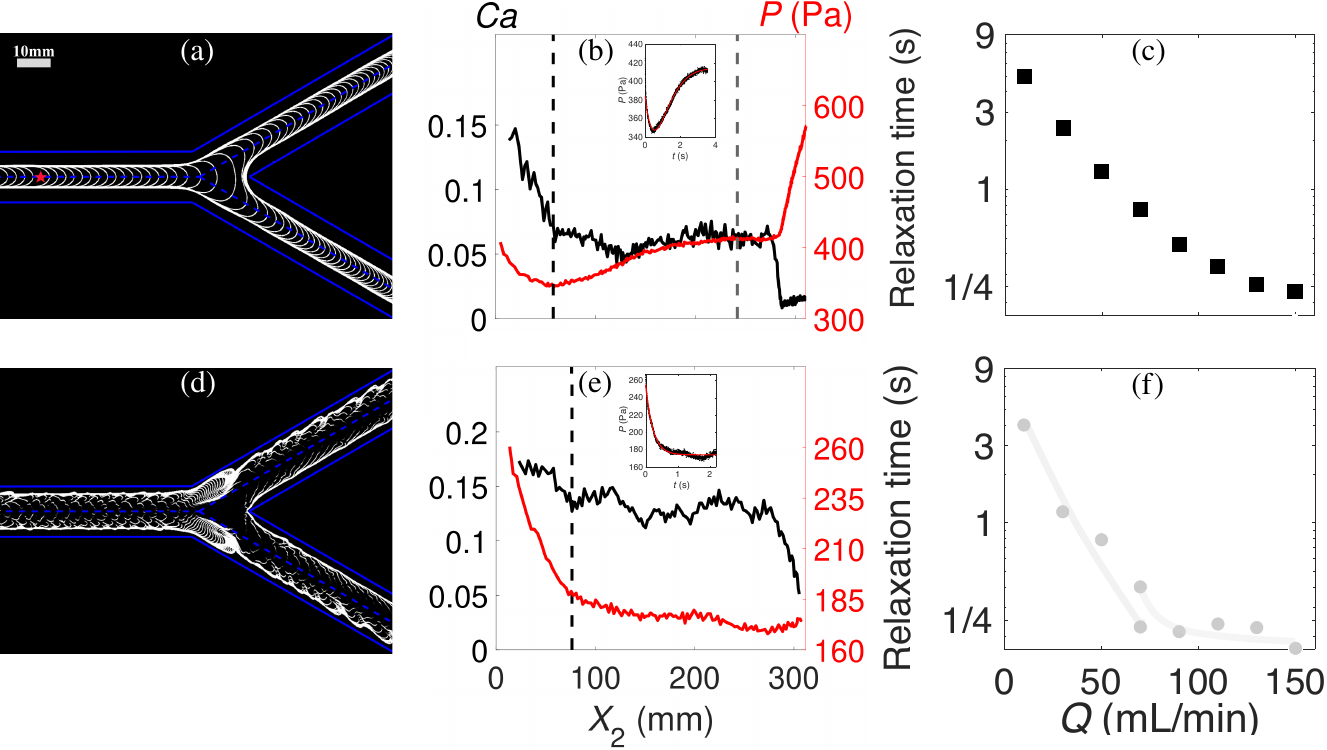}
	\caption{
          Spatiotemporal composite image for (a) $Q$ = 70~ml/min and $A_{i}$ = 0.91/0.89 in the main/daughter channel and (d)~$Q$ = 110~ml/min and $A_{i}$ = 0.36/0.32 in the  main/daughter channel. The time interval between sequential contours is 0.017~s. (b,~e)~Capillary number $Ca$ and finger pressure $P$ as a function of scaled finger tip position in the daughter channel $X_{2}/W$ for experiments in~(a) and~(d), respectively. $Ca$ and $P$ are approximately constant beyond the vertical dashed lines until the air finger reaches the end of one of the daughter channels (for~(b), $\overline{Ca}$ and $\overline{P}$ are reached at different locations, $X_{2}/W = 4.6$ and 15.5, respectively). Insets show post-bifurcation time-evolution of the finger pressure $P$ in the same experiments (black dots) and fits to the data (red lines) assuming (b)~exponential and (e)~modulated exponential decay to a constant pressure, see~(\ref{eq:3.1})-(\ref{eq:3.2}), respectively.
 (c, f)~The relaxation time, obtained from fits such as in (b, e),  as a function of the injection rate on a semi-log scale obtained for experiments with (c)~$A_{i}$ = 0.89 and (f)~$A_{i}$ = 0.32 in the daughter chanel. The squares and circles were obtained using~(\ref{eq:3.1}) and~(\ref{eq:3.2}), respectively. Lines in (f)~are drawn to guide the eye. 
	}
	\label{fig:PFitting}
      \end{figure}
      
We now turn to the study of the transient evolution of the daughter fingers that emerge from the bifurcation region. The flow rate in each of the daughter channels is half of the injection flow rate, and the fingers propagate into channels that are marginally less stiff and therefore more collapsed than the main channel (see section~\ref{ExpMethods}). Moreover, the membrane is marginally less collapsed near the outer boundary at the inlet of the daughter channels due to the geometry of the network and the orientation of the imposed pre-stress, see figure~\ref{fig:Ai}(a).

Having determined that the reopening of the bifurcation region depends crucially on whether viscous or elastic forces dominate, analogously to straight channels, we now proceed to examine the recovery towards steady propagation in these two limits.
Figure~\ref{fig:PFitting} compares finger propagation in the viscous regime ($Q=70$~ml/min and $A_i=0.91$, i.e., relatively low injection rate and low initial collapse) and in the elastic regime ($Q=110$~ml/min and $A_i=0.36$, i.e., relatively high injection rate and high initial collapse). 
In the viscous regime~(figure~\ref{fig:PFitting}(a)), the finger is symmetric about the centreline of the main channel and
the apex of the bifurcation, where its tip divides. However, the daughter fingers are initially slighly asymmetric in line with the slight asymmetry of the daughter channels. They recover a steadily-propagating symmetric shape as they propagate downstream, with approximately constant $\overline{P}$ and $\overline{Ca}$, which are both lower than the values in the main channel. 
In contrast, the flattened finger tips in the elastic regime (figure~\ref{fig:PFitting}(d)) penetrate the most collapsed parts of the channel as fingers transition from one feathered pattern in the main channel to a different one in the daughter channels.

For the experiments in figure~\ref{fig:PFitting}(a), the recovery of a steady state involves the monotonic decay of $Ca$ to a constant value $\overline{Ca}$, which is less than half of its value in the main channel (figure~\ref{fig:PFitting}(b)). However, $P$ evolves non monotonically to reach its constant value $\overline{P}$ . Thus, values for the average capillary number $\overline{Ca}$ plotted in figure~\ref{fig:CaP_main}(b) were obtained from experimental data by averaging $Ca$ once it reached a constant value and errors were estimated as the standard deviation of the data within an experiment. In order to extract $\overline{P}$ from the time-evolution of the pressure data, also shown in figure~\ref{fig:PFitting}(b), we fitted an exponentially decaying function of the form 
\begin{equation}\label{eq:3.1}
P(t)=C_{1}\exp(-t/T)\cos(C_{2}t+C_{3})+\overline{P},
\end{equation}
where $C_{i}$ ($i=\overline{1, 2,3}$) and $T$ are fitting parameters. The fit is shown in the inset to figure~\ref{fig:CaP_main}(b). When fitting, we considered a range of possible $\overline{P}$ and selected the value which provided the best least square fit of a straight line to $\ln\frac{P(t)-\overline{P}}{\cos(C_{2}t+C_{3})}$. The fit also produced a characteristic time scale of transient decay $T$, which we refer to as the relaxation time.
Figure~\ref{fig:PFitting}(c) indicates that the relaxation time decreases monotonically with flow rate for viscous reopening.

For the experiments in figure~\ref{fig:PFitting}(d), both $Ca$ and $P$ decay monotonically to their constant values $\overline{Ca}$ and $\overline{Pa}$, respectively, at approximately the same position $X_{2}/W$ = 5 (figure~\ref{fig:PFitting}(e)).
For this type of transients, exponentially decaying functions of the form 
\begin{equation}\label{eq:3.2}
P(t)=C_{0}\exp(-t/T)+\overline{P},
\end{equation}
were fitted to the time-evolution of the pressure data in order to extract $\overline{P}$; see the inset to figure~\ref{fig:PFitting}(e). As before, $C_{0}$ and $T$ are fitting parameters.
For this higher level of collapse, the relaxation time shown in figure~\ref{fig:PFitting}(f) is discontinuous with flow rate. There are two groups of points corresponding to the data on either side of the vertical dashed line in figure~\ref{fig:CaP_main}(a). This is related to the fact that the dominance of viscous or elastic forces can vary throughout the channel for intermediate values of $Q$ and $A_i$, as demonstrated in \S~\ref{Section3.3.1}, and this in turn affects the length of transients.

\begin{figure}
  \centering
 \includegraphics[width=0.9\linewidth]{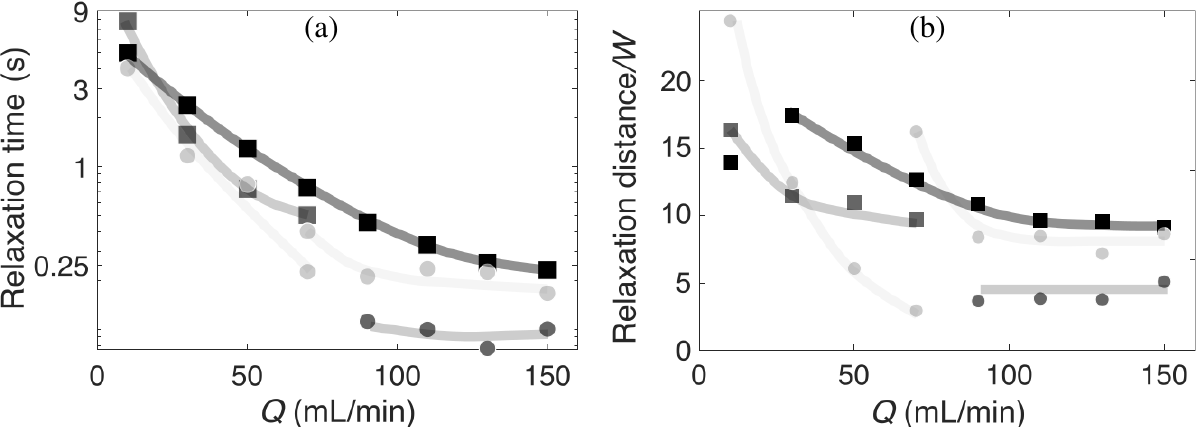}
\caption{Relaxation (a) time and (b) distance as a function of $Q$ for $A_{i}$ = 0.91/0.88 (black markers),  $A_{i}$ = 0.64/0.57 (grey markers) and $A_{i}$ = 0.36/0.32 (light grey markers) in the main/daughter channel. As in figure~\ref{fig:PFitting}, the squares and circles were obtained using~(\ref{eq:3.1}) and~(\ref{eq:3.2}), respectively, and lines are drawn to guide the eye.}
\label{fig:scales}
\end{figure}

Figure~\ref{fig:scales}(a) shows similar variation of the relaxation time with injection rate for all experiments performed with different levels of initial collapse.
The shortest timescales of reopening, shown with dark grey circles in figure~\ref{fig:scales}(a), are for fingers that adopt asymmetric shapes throughout the Y-channel, including in the bifurcation region, e.g. for experiments with $A_i=0.64/0.57$ in the main/daughter channel. This is consistent with the fact that weakly asymmetric fingers are formed in the bifurcation and thus, adjustment to steady asymmetric shapes is likely to be rapid.

A more practical description of the transient dynamics is in terms of a relaxation distance. We define it as the length of daughter channel required for the finger pressure to decay to within 2\% of its value $\overline{P}$, a criterion which allows us to account for small fluctuations in $\overline{P}$. Results shown in figure~\ref{fig:scales}(b) confirm the trends obtained for the relaxation time in figure~\ref{fig:scales}(a) but the disconnection between data associated with different modes of propagation is amplified. For example, it is clear that at high initial collapse,  the relaxation distance follows two separate trends with $Q$, resulting in an overall non-monotonic variation with $Q$. At smaller values of $Q$, the lengthscale of decay routinely requires a long daughter channel exceeding ten channel widths. The fact that the relaxation distance reaches a constant value for high $Q$ suggests that it may be possible to build tractable multiscale models of elastic-walled networks based on the assumption of steady reopening in this regime.

\section{Conclusion}
\label{Conclusion}

In this paper we have presented a bench top model of the reopening of a liquid-filled, collapsed branching airway via the propagation of an air finger. Our model consists of a straight elasto-rigid Hele-Shaw channel which bifurcates into two daughter channels of similar geometry. We find that the influence of the channel bifurcation on the reopening dynamics depends significantly on the level of initial collapse and the rate of air injection. At lower levels of collapse, following the decay of transients, steady modes of finger propagation are recovered downstream of the bifurcation, which are similar to those in the main channel for suitably reduced flow rates. In this regime, the dynamics are insensitive to small differences in channel geometry between main and daughter channels.

However, at high levels of initial collapse, the finger dynamics becomes more complex and can exhibit sensitivity to small changes to the initial collapse, as evidenced by qualitatively  different reopening processes in the main and daughter channels where initial collapse is only slightly larger. This sensitivity concurs with the results of \citet{Cuttle2020} and \citet{Thesis_Fontana2021} and in the bifurcating channel, it can lead to multiple reopening scenarios downstream of the bifurcation despite apparently identical fingers in the main channel.  

We find that the recovery of a steady state downstream of the bifurcation is regulated by the same reopening mechanics that underpin finger propagation in straight elasto-rigid channels. At smaller flow rates and/or lower levels of initial collapse, it is dominated by the balance between viscous and surface tension forces, and we find transients that persist for longer and over a large distance. As the flow rate and/or level of initial collapse decrease, the system gradually transitions to the regime in which reopening is controlled by a balance between surface tension and elastic forces, i.e. the elastic regime. This is associated with an overall decrease in time scale and length scales of transients, but the route is complex because experiments can feature different reopening regimes in the main channel, bifurcation and daughter channels for intermediate parameter values. In the elastic regime, the time- and length- scales of the steady state recovery are the shortest and reach a constant plateau. 

A key assumption in all tractable models of multiphase flows in elasto-rigid networks is that their reopening is steady. Our experimental observations confirm that in channels of moderate length ($\lesssim$ ten channel widths) this assumption is reasonable for a limited range of parameters in the elastic regime, where transients are relatively short-lived. However, we also demonstrate complexity related to dynamical sensitivtity which is not accounted for in existing theoretical models, and thus our findings may be used to inform future modelling assumptions. 

\begin{acknowledgements}
The authors would like to acknowledge Martin Quinn for his technical support. The work was funded by the EPSRC [Grant No. EP/R045364/1]. H. L. would like to thank the China Scholarship Council for supporting him; F.B. would like to acknowledge the Royal Society (Grant No. URF/R1/211730). For the purpose of open access, the authors have applied a Creative Commons Attribution (CC BY) licence to any Author Accepted Manuscript version arising.
\end{acknowledgements}
    
\appendix

\section{}\label{appA}

\begin{figure}
  \centerline{\includegraphics[width=\linewidth]{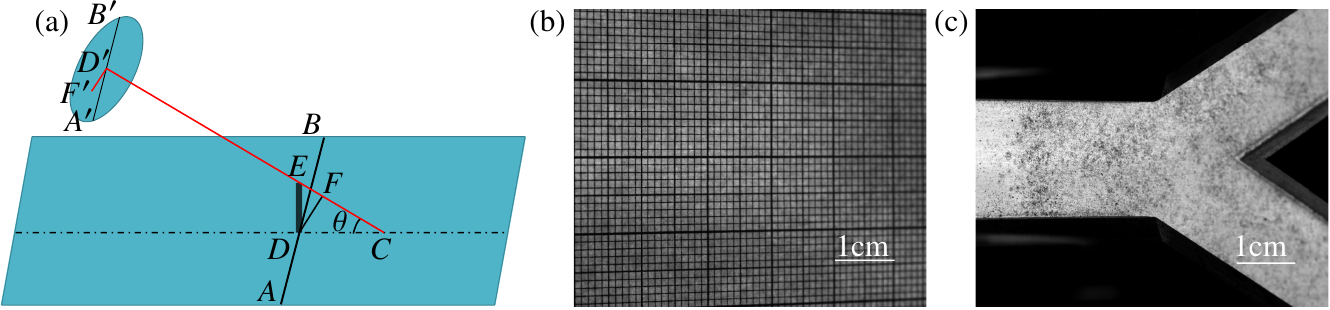}}
  \caption{(a) Schematic diagram of the principle behind detecting line deformations. Examples of (b) the calibration and (c) the reference
image.}
\label{fig:app1}
\end{figure}

The method for reconstructing the vertical deflection of the membrane in our experiments relies on the method for reconstructing the vertical line displacement used by \citet{Lister2013, Ducloue2017b, PihlerPuzovic2015, Cuttle2020}. We start by considering a horizontal plane $ABC$, containing a point $D$ that lies in the middle of the line $AB$, see figure~\ref{fig:app1}(a). To reconstruct its position as it moves vertically to the point $E$, we place a camera at an angle  $\theta$ with the horizontal. This implies that the image plane $A^{\prime}B^{\prime}D^{\prime}F^{\prime}$ is parallel with the plane $ABDF$, where $FD'$ is at the angle $\theta$ with the plane $ABC$  and  $DF$ forms an angle $\theta$ with the vertical displacement $DE$.
Thus, it is possible to find the true displacement of the point $D$, $DE$, from the in-image displacement $D'F'$ using the conversion ratio $\mathcal{C} = {DE}/{D'F'} = {DE}/{DF}\cdot{DF}/{D'F'} ={DF}/{D'F'}\cdot{1}/{\cos{\theta}} = {\mathcal{S}}/{\cos{\theta}}$, where $\mathcal{S}$ is a calibration factor, which, in principle varies for different points on the line $AB$. However, as the width of the viewing area is much smaller than the camera-object distance in our experiments, we assume that the scale factor $\mathcal{S}$ stays constant along the line $AB$, and obtain it by relating the number of the in-image pixels to the known length of the line $AB$. This allows us to calculate a vector of vertical displacements $V_{VD}$ of points on the line $AB$ from a vector of the corresponding in-image displacements $V_{ID}$ using the relationship $V_{VD} = V_{ID} \cdot \mathcal{C}$. 

To reconstruct the vertical deformation of the whole membrane, we needed to capture the deflection of points lying on many parallel lines which cover the region of interest. Before the channel was filled with liquid and collapsed to be subsequently inflated, a millimetric tracing paper was laid flat on the flat elastic wall and a calibration image was taken (see figure~\ref{fig:app1}(b)). For each of the parallel lines, the scale factor $\mathcal{S}$ and the angle $\theta$ were found from the calibration image, and recorded in vectors $V_{\mathcal{S}}$ and $V_{\theta}$, respectively. Thus, the conversion ratio vector $V_{\mathcal{C}}=\{V^i_{\mathcal{C}}\}$ for all lines was obtained using the formula $V^i_{\mathcal{C}}=V^i_{\mathcal{S}}/\cos(V^i_{\theta})$. Then by knowing the in-image displacements $M_{ID}$ of different points on the membrane, we could reconstruct the vertical deformation field $M_{VD}$ using the matrix relationship 
\begin{equation}
M_{VD}=M_{ID}\cdot V_{\mathcal{C}}.\label{mat_DIC}
\end{equation}

In our experiments, membrane deformations comprised both in-plane as well as out-of-plane displacements. Both types of displacements are typically reconstructed using stereo or three-dimensional digital image correlation with two synchronised digital cameras, or using single-camera stereo-digital image correlation techniques with additional optical assistance~\citep{Pan2018}. These techniques track three dimensional positions of random particles, deposited on the deforming object. We also traced a random speckle pattern on the elastic membrane created by sprinkling food powder onto it while it was still flat, as shown in the reference image figure~\ref{fig:app1}(c). However, we chose to neglect the deformations due to in-plane stretching, assuming that the membranes predominantly deformed through vertical displacement. This allowed us to employ one camera as in experiments with a single line, and apply a two-dimensional digital image correlation technique (2D DIC) to obtain the in-image displacements $M_{ID}$ of the speckle pattern. We used the 2D DIC method available as an open source MATLAB software Ncorr v1.2 \footnote{The software is available at {\tt http://www.ncorr.com}.}, which employs a sum-squared difference correlation criteria~\citep{Tong2005} to match the particles before and after the deformation, and the B-spline interpolation~\citep{Schreier2000} to enhance the method for finding the displacement field to sub-pixel accuracy. We found that imaging the membrane using a CMOS camera (Teledyne Dalsa CR-GM00-H1400, resolution 1400 $\times$1024 pixels) at 1 frame per second during the collapse and at 60 frames per second during the reopening experiments, respectively, was sufficient to reproduce reliable in-image displacement measurements which could then be used to extract the vertical deformations using (\ref{mat_DIC}). 

\begin{figure}
  \centerline{\includegraphics[width=\linewidth]{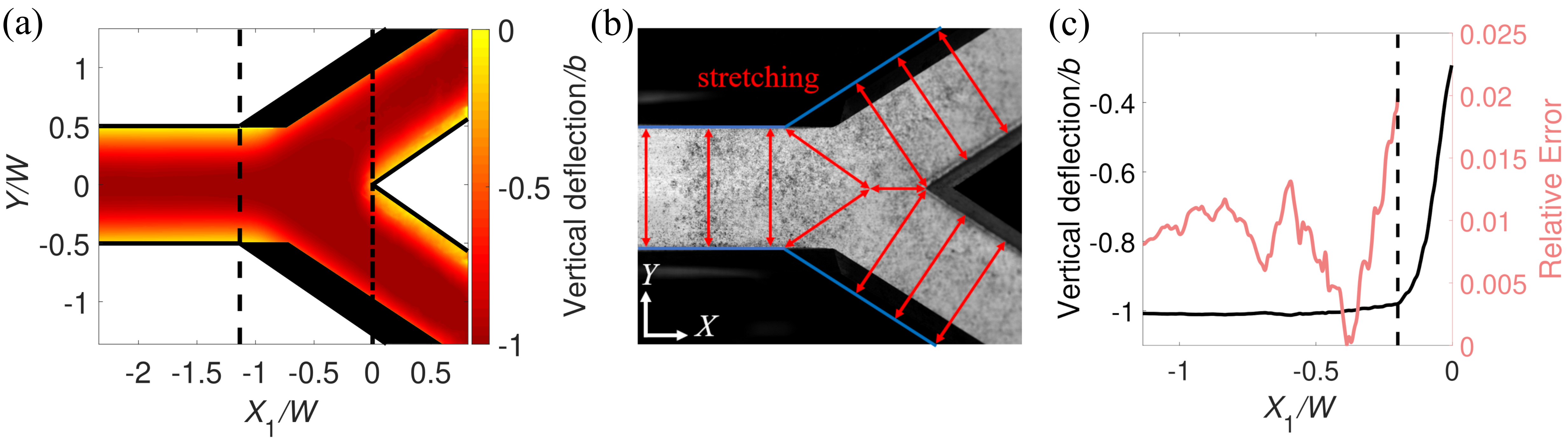}}
  \caption{(a) Membrane profile for $A_{i}$ = 0.33 before the start of air injection with colour-coding corresponding to the vertical deformation of the membrane scaled by $b$. Regions of the channel, where the vertical deformation could not be reconstructed, are black. Lines denote the boundaries of the channel (solid lines), the start of bifurcation region (vertical dashed line) and the apex of the bifurcation (vertical dash-dotted line). (b) Illustration of principal stretching directions in different regions of the Y-channel, whose boundaries are shown using (blue) solid lines. (c) Recorded membrane deformation in the longitudinal mid-plane of the bifurcation region (left axis) and the estimated relative error of this measurement (right axis). The vertical dashed line corresponds to $X_1/W=-0.2$, beyond which the relative error can not be estimated because there is no opposite wall contact.}
\label{fig:app2}
\end{figure}

When reconstructing the vertical displacements $M_{VD}$ of the membrane, e.g., in figure~\ref{fig:app2}(a), we only used the in-image displacements in the flow direction (marked as $X$ in the image figure~\ref{fig:app2}(b)), and while the horizontal in-plane stretching of the membrane was much smaller than its vertical displacement, it was nevertheless the primary error source for the proposed method over the majority of the channel area. The predominant direction of the stretching, schematically shown with red double arrows in figure~\ref{fig:app2}(b), was different in the main channel, the bifurcation region and the daughter channels. In the main channel, the reconstructed deflections were not affected significantly by the in-plane displacement which was mostly confined to the $Y$-direction. However, in the bifurcation region and the daughter channels, the stretching contributed significantly to the $X$-directional in-image displacements. This was especially true at the apex of the bifurcation, where the direction of stretching coincided exactly with the $X$-direction. Conversely, in the main channel the error was expected to be the largest near the side boundaries, which are not captured precisely by the camera due to a small unavoidable misalignment between the channel and the metal frame clamping the elastic membrane. 

In order to quantify these errors of the measurement method, we utilised the deformation results obtained for the cross-sectional area of $A_{i}$ = 0.33 in the main channel, when the opposite wall contact occurred centrally in all parts of channel (see figure~\ref{fig:app2}(a)). By assuming the real vertical displacements along the central lines and the boundaries of the main/daughter channels to be -0.5~mm and 0~mm, respectively, we extracted the relative errors on the vertical displacement in these regions by comparing them to the depth of undeformed channel, as done in the bifurcation region in figure~\ref{fig:app2}(c). This yields errors of up to $1.3\%$ and $3.9\%$ on the deflection along the centrelines of the main and the daughter channels, respectively. The relative error near the boundaries of the channels increases to $7\%$ and $16.7\%$, respectively. 
The greatest relative error of $33\%$ was estimated at the bifurcation apex. However, we are primarily interested in the vertical displacements of the channel centreline in the region of $X_{1}/W \leq -0.2$ (as discussed in \S 3.2), where the relative error on the displacement is less than $2\%$ in figure~\ref{fig:app2}(c). For $-0.2 \leq X_{1}/W < 0$ in figure~\ref{fig:app2}(c), we could not estimate the errors using the arguments above as this region corresponds to where the membrane adjusts its deformation from the opposite wall contact to being clamped at the apex.

\bibliography{references}

\end{document}